\documentclass[aps,prl,showpacs,preprint,amsfonts,amssymb,amsmath]{revtex4}

\usepackage{graphicx}

\begin{document}

\title{{\it Ab initio} analysis of electron-phonon coupling in molecular 
devices}

\author{N. Sergueev, D. Roubtsov, and Hong Guo}
\address{Center for the Physics of Materials and Department of Physics, 
McGill University, Montreal, PQ, Canada, H3A 2T8}

\begin{abstract}
We report first principles analysis of electron-phonon coupling in molecular 
devices under external bias voltage and during current flow. Our theory and 
computational framework are based carrying out density functional theory within 
the Keldysh nonequilibrium Green's function formalism. We analyze which
molecular vibrational modes are most relevant to charge transport under 
nonequilibrium conditions. For a molecular tunnel junction of a 
1,4-benzenedithiolate molecule contacted by two leads, the low-lying modes 
of the vibration are found to be most important. As a function of bias voltage,
the electron-phonon coupling strength can change drastically while the 
vibrational spectrum changes at a few percent level.
\end{abstract}

\maketitle

\begin{figure}
\begin{center}
\leavevmode
\includegraphics[width=10cm]{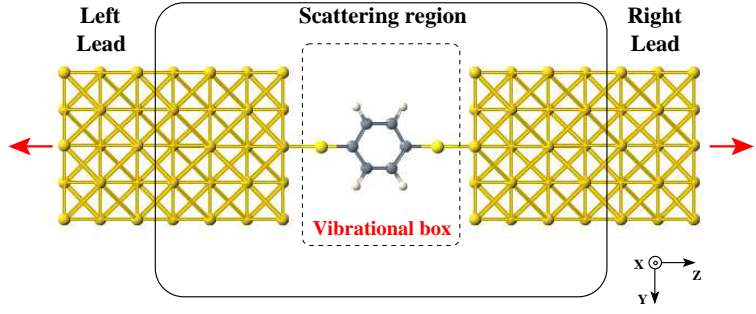}
\end{center}
\caption{Schematic plot of a Metal-BDT-Metal molecular 
tunneling junction (${\rm BDT}={\rm C}_{6}\,{\rm H}_{4}\,{\rm S}_{2}$). 
The electrodes consist of repeated unit cells extending to 
$z=\pm\infty$. The scattering region contains several layers of electrodes 
and the molecule. The vibrational box lies inside the scattering region. 
}
\label{fig1}
\end{figure}

One of the most important questions concerning charge transport in molecular 
electronic devices is the role of electron-phonon (e-p) interaction. Here, 
``phonon'' refers to quantized molecular vibrational modes which couple to 
various scattering states of the device. A typical molecular device has the
Metal-Molecule-Metal (MMM) configuration schematically shown in Fig. \ref{fig1},
where metal leads extend to far away and bias voltages can be applied so 
that a current flows through. The problem of predicting vibrational 
spectra and e-p coupling strength for such an open system during current flow, 
within self-consistent first principles including all atomic details of the 
molecule as well as the leads, is a serious theoretical challenge that has 
not been satisfactorily addressed. A particularly important problem is to 
understand which vibrational mode couples to which scattering state at what 
bias voltage \cite{lorente2000,montgomery2003a}. It is the purpose of 
this article to address this issue.

Experimentally, single molecule vibrational spectra can be measured by 
inelastic tunneling spectroscopy (IETS) \cite{ho2002,Reed2004,Mayer05}. 
Theoretically, various models have been applied to understand IETS and to 
investigate effects of e-p interaction based on tight binding atomistic 
Hamiltonians \cite{montgomery2003,ratner2004,asai2004}. 
Recently, Frederiksen 
{\it et al.} \cite{jauho2004} reported a first principles analysis of inelastic 
current due to e-p interactions in an Au chain, in which the relevant 
vibrations are along the chain length. In their theory \cite{jauho2004}, 
the vibrational spectra was obtained using a plane-wave basis density 
functional theory (DFT) code in a cluster configuration at equilibrium, and 
the dynamic matrix was evaluated using a finite differencing scheme. Transport 
properties were then obtained using the Transiesta package \cite{transiesta} 
with LCAO basis, and e-p scattering was included at the level of 
self-consistent Born approximation. 

In order to investigate voltage dependence of the e-p interaction in MMM
devices, however, quantized molecular vibrations and electrons need to be 
treated on equal footing at {\it nonequilibrium}. We accomplish this by 
carrying out DFT atomic analysis within the Keldysh nonequilibrium Green's 
function (NEGF) formalism \cite{mcdcal}. In addition, we calculate the dynamic 
matrix within the NEGF-DFT formalism \cite{mcdcal} by evaluating analytical 
formula rather than numerical finite differencing \cite{jauho2004}: this is 
more general and more accurate so that all the phonon modes and e-p couplings 
can be obtained for complicated systems. For a molecular tunnel junction of a 
1,4-benzenedithiolate (BDT) molecule contacted by two metallic electrodes 
(see Fig. \ref{fig1}), we found that the low-lying modes of the vibration are the 
most important for e-p coupling. As a function of bias voltage, the coupling 
strength can change drastically while the vibrational spectrum changes at a 
few percent level. 

We start from the NEGF-DFT formalism documented in Ref. \onlinecite{mcdcal},
in which the density matrix $\hat{\rho}$ is calculated by NEGF $G^<$, 
$\hat{\rho}\sim \int dE\, G^<(E)$. This way, we naturally take into account
external bias voltage and open device transport boundary condition. 
After the Kohn-Sham Hamiltonian $\hat{H}_{KS}[\,\hat{\rho}\,]$ of the device
is obtained self-consistently, the total energy of the scattering region 
(see Fig. \ref{fig1}), as well as all transport properties of the device, can be 
obtained \cite{mcdcal}. Importantly, the NEGF-DFT formalism allows one to 
obtain $\hat{H}_{KS}$ and total energy $E(\{{\bf R}_i\}, V_b)$ as 
functions of external bias $V_b$, (${\bf R}_i$ is the position of the 
$i$-th atom).

The phonon (vibrational) eigenvectors ${\bf e}_{\nu}$ and frequencies 
$\omega_{\nu}$ ($\nu$ is the mode index) are obtained by diagonalizing the 
dynamic matrix (Hessian matrix),
\begin{equation}
{\cal H}_{j,j'}=\nabla_{{\bf R}_j} \nabla_{{\bf R}_{j'}} 
E(\{ {\bf R}_{i} \},V_b)/\sqrt{M_j M_{j'}}, 
\end{equation}
where $M$ is the mass of an
atom. Once $({\bf e}_{\nu},\,\omega_{\nu})$ are obtained, the e-p interaction 
strength $g^{\nu}$, defined by the e-p Hamiltonian \cite{gunnarsson}, can be 
obtained from the standard expression:
\begin{equation}
g^{\nu}_{j,\mu;\,j',\mu'} = \sum_i\sqrt{\frac{\hbar}{2M_i\omega_{\nu}}}
\,{\bf e}(\nu,i) \langle\phi_{j,\mu}\vert\,\nabla_{{\bf R}_i} 
\hat{H}_{KS}\,\big\vert \phi_{j',\mu'}\rangle . 
\end{equation}
Here,
$\bigl\vert \phi_{j,\mu} \bigr \rangle $ is a basis function for 
orbital $\mu$ ($\mu = s,p,d$) of the $j$-th atom. The derivative of the KS 
Hamiltonian is carried out by fixing the atomic positions at their stationary 
locations ${\bf R}_i^0$. (${\bf R}_i^0$ is where the $i$-th atom feels no 
force \cite{foot1}.) Although the Hessian matrix ${\cal H}_{j,j'}$ and the e-p coupling matrix
$g^{\nu}_{j,\mu;\,j',\mu'}$ appear to have the same form as those for 
equilibrium system, the derivatives of $\hat{H}_{KS}(\{{\bf R}_i\}, V_b)$ and 
$E(\{{\bf R}_i\}, V_b)$ with respect to ${\bf R}_i$ propagate to derivatives 
of NEGF $G^<(E)$ that includes nonequilibrium physics \cite{foot1}.

The e-p coupling is characterized by a dimensionless parameter \cite{gunnarsson} 
$\lambda_{e-p}$ which is contributed by all phonon modes,
\begin{equation}
\lambda_{e-p}=\sum_{\nu} \lambda_{\nu}, \,\,\,\,\,\,
\lambda_{\nu}\equiv {DOS}(\varepsilon_{\rm F}^{})
\frac{\big\vert \langle\,g^{\nu}\rangle 
\big\vert^{2}}{\hbar\omega_{\nu}} .
\label{lambda0}
\end{equation}
Here, $DOS(E)$ is the density of states of the scattering region and 
$\varepsilon_{\rm F}^{}$ the Fermi energy of the leads. As we are interested 
in e-p coupling for quantum transport, the $g^{\nu}$ matrix is averaged over 
scattering states $\Psi_{\rm sc}$, which are obtained by the NEGF-DFT numerical 
package \cite{mcdcal} for any MMM device:
\begin{equation}
\langle g^{\nu}(E,E')\rangle=\langle\Psi_{\rm sc}(E)\vert \,g^{\nu}\,\vert
\Psi_{\rm sc}(E')\rangle .
\label{gnu1}
\end{equation}
If both scattering states have the same energy $E=E'=\varepsilon_{\rm F}^{}$, 
we call such an e-p coupling the ``elastic'' one, $\lambda_{\nu}^{{el}}$. When 
$E=\varepsilon_{\rm F}^{}$ and $E'=\varepsilon_{\rm F}^{}\pm \hbar\omega_{\nu}$,
we call it the ``inelastic'' $\lambda_{\nu}^{\pm}$. Finally, when the MMM 
device is under an external bias voltage $V_b$, we further average 
$\lambda_{\nu}$ over the transport energy window $(\mu_{L}^{},\,\mu_{R}^{})$ 
where $\mu_{R/L}^{}$ are the electrochemical potentials of the right/left leads
and $|\mu_{R}^{}-\mu_{L}^{}|=eV_b$. Hence, at nonequilibrium, we obtain
\begin{equation}
\lambda_{\nu}^{el}(V_b)
=\int_{\mu_{L}^{}}^{\mu_{R}^{}} \frac{dE}{eV_{b}}\,
{ DOS}(E)\,\frac{\big\vert \langle\,g^{\nu}(E,E) \rangle \big\vert^{2} }
{\hbar\omega_{\nu}} .
\label{lambda1}
\end{equation}
The inelastic coupling $\lambda_{\nu}^{\pm}(V_{b})$ is calculated by replacing 
$g^{\nu}(E,E)$ with $g^{\nu}(E,\,E\pm\hbar\omega_{\nu})$ in Eq. (\ref{lambda1}).

The above theoretical formalism is implemented into our NEGF-DFT package
McDCAL \cite{mcdcal}. In numerical calculations, we further define a
``vibrational box'' inside the MMM device (see Fig. \ref{fig1}) which contains 
the atoms of interest. Typically, the vibrational box include the molecule and 
perhaps a few layers of the nearest lead atoms. The atomic indexes in the above 
formalism refer to those inside the vibrational box. 

In the following, we investigate general features of vibrational spectra and 
e-p coupling during nonequilibrium transport using the model MMM device of 
Fig. \ref{fig1}, {\it i.e.},  a BDT molecular wire \cite{foot2}. For each bias 
voltage, we iterate the KS Hamiltonian of the device to numerical convergence 
using the NEGF-DFT method \cite{mcdcal}; the atomic positions in the scattering
region must also be relaxed for each applied bias. Afterward, the vibrational 
spectrum and the e-p coupling are obtained. As a check, we calculated 
$\omega_{\nu}$ of an isolated BDT using our NEGF-DFT formalism and obtained 
reasonable agreement, to within $\le 5-6$~ \%, with experimental data collected 
by Raman spectroscopy and other methods \cite{Raman1}. We also checked that 
the {\it diagonal} matrix elements of the e-p coupling are non-zero for modes 
having the ${A}_{\rm g}$ symmetry and are zero for other modes, in agreement 
with selection rules from group theory \cite{GTB}.

When the BDT is placed between the leads (see Fig. ~\ref{fig1}), new properties 
arise. First, several low-lying modes that do not exist for isolated BDT are 
found to play important roles. These modes include the center-of-mass and 
libration ({\it CM}$\,(i)$ and {\it LB}$\,(i)$, $i =X,Y,Z$) with energy 
$\hbar\omega_{\nu} \simeq 8-14$ meV. Clearly, the presence of leads breaks the 
translational and rotational symmetries and produces these low-lying modes. 
Second, many vibrational frequencies are renormalized, to $\simeq 10-30$ \%,
from that of the isolated molecule. This is especially true for modes with 
strong sulphur oscillations in the BDT. For large bias, $V_{b} \approx 1$ V, 
we found that $\omega_{\nu}$ changes up to 2 \% while ${\bf e}_{\nu}$ changes 
up to 5 \% compared with the vibrational spectrum at $V_b=0$. 
At $V_b=0$, it turns out that all the modes can 
be classified by the same $D_{2h}$ point group as in the case of an isolated BDT; 
at $V_b\neq 0$, they can be classified by the $C_{2v}$ point group. 

\begin{figure}[t]
\begin{center}
\leavevmode
\includegraphics[width=12cm]{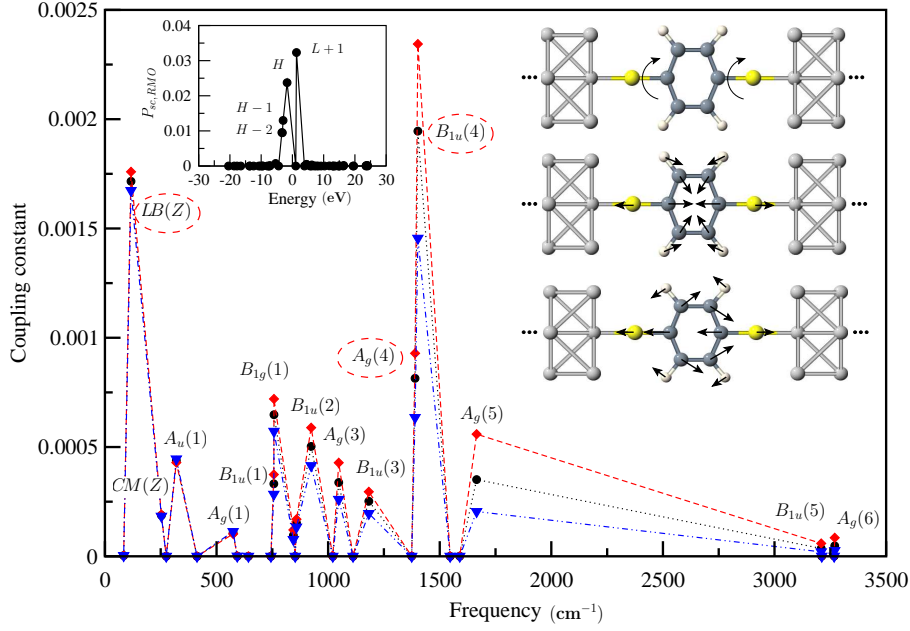}
\end{center}
\caption{ 
Dimensionless e-p coupling constants $\lambda_{\nu}^{el}$ (circles), 
$\lambda_{\nu}^{-}$ (triangles), $\lambda_{\nu}^{+}$ (diamonds) as a function 
of vibrational frequency $\omega_{\nu}$ at $V_b=0$. The lines are guide to 
the eye. The modes are classified using the $D_{2h}$ point group. The left 
inset shows projection $P_{sc,\,RMO}$ of a scattering state at Fermi energy 
(shifted to zero) onto molecular orbitals (RMO) of the BDT. Letters $H$ and 
$L$ mean HOMO and LUMO. In the right side insets, we show eigenvectors of 
several important eigenmodes of the BDT at small biases, namely, the libration 
$LB(Z)$ (top), $A_{\rm g}(4)$ (middle), and $B_{\rm 1u}(4)$ (bottom).
}
\label{fig2}
\end{figure}

Fig. \ref{fig2} plots the e-p coupling $\lambda_{\nu}$ versus 
$\hbar\omega_{\nu}$ at $V_b=0$. For small bias voltages, $V_b \le 0.5$~ V, 
$\lambda_{\nu}$ does not change qualitatively. Most clearly shown is that
some phonon modes give distinctly larger e-p coupling to scattering states 
than others (see Eq. (\ref{gnu1})). Beside the expected in-plane 
${A}_{\rm g}(n)$ modes, modes of other symmetries are also responsible for 
the peaks in $\lambda_{\nu}$ (see the right insets of Fig. \ref{fig2} for a 
few important modes). Notable are the in-plane modes ${B}_{\rm 1u}(n)$, the 
center-of-mass mode $CM(Z)$, the out-of-plane modes ${A}_{\rm u}(n)$, 
${B}_{\rm 1g}(1)$, and the libration $LB(Z)$. Recall that for vibrational 
spectroscopy on free BDT such as Raman or infrared, there are always selection 
rules of modes \cite{Raman1}. For a BDT device, however, our results suggest 
that no obvious selection rules are followed because many modes with very 
different symmetries manifest.  Interestingly, among the low-lying modes with 
$\omega_{\nu}<1000$ cm$^{-1}$ ($\hbar\omega_{\nu}<0.12$ eV), the ``breathing'' 
modes ${A}_{\rm g}(1)$ and ${A}_{\rm g}(2)$ are not the most important ones  
for coupling to scattering states ($\lambda_{\nu}\simeq 1-2 \times 10^{-4}$), 
although these modes are important for a free BDT. For the BDT device, the 
total e-p coupling (see Eq. (\ref{lambda0})) is \,$\lambda_{e-p}^{el} \approx 
\lambda_{e-p}^-\approx 7 \times 10^{-3}$ at $V_b=0$.

To understand why modes with symmetry other than $A_{\rm g}$ can couple 
to scattering states, as shown by some of the peaks in Fig. \ref{fig2}, we 
consider Eq. (\ref{gnu1}). For a {\it free} BDT, the electronic wave function in 
Eq. (\ref{gnu1}) is a molecular orbital, and for each orbital one obtains a 
value $\langle g^\nu \rangle$. Hence for a free BDT the relevant e-p coupling 
is a diagonal matrix in orbital space: only those vibrational modes with 
$A_{\rm g}$ symmetry give nonzero values to these diagonal matrix 
elements \cite{GTB}. For transport, however, the wave function appearing in 
Eq.~ (\ref{gnu1}) is a scattering state which is roughly a linear combination 
of many molecular orbitals. Therefore it is possible to have off-diagonal 
matrix elements in the coupling matrix so that modes with symmetries other 
than $A_{\rm g}$ can also contribute. This can be substantiated as follows. 
We project scattering states $\Psi_{\rm sc}(E)$ onto ``renormalized molecular 
orbitals'' (RMO) of the BDT in the MMM device \cite{Larade}. 
The RMO's are 
obtained by diagonalizing the Hamiltonian sub-matrix that corresponds to the 
BDT molecule, and this sub-matrix is a part of the total KS Hamiltonian of the 
entire MMM device \cite{Larade}. Note that RMO's can be different from the 
original molecular orbitals of an isolated BDT due to charge transfer from the 
leads to the molecule and external bias potentials. The projection is 
characterized by the quantity $P_{sc,\,RMO}\equiv \vert\langle
\Psi_{\rm sc}(\varepsilon_{\rm F}^{})\,\vert\, RMO\rangle\vert ^{2}$ plotted in
the left inset of Fig. \ref{fig2} as a function of energy. 
We found that all 
scattering states $\Psi_{\rm sc}$ near the Fermi energy are contributed by 
the {\it same} {\it several} dominant RMO's at low bias. Therefore one can 
well consider that $\Psi_{\rm sc}$ is a linear combination of these few RMO's 
and $g^{\nu}$ of Eq. (\ref{gnu1}) is contributed mostly by them, {\it i.e.}, 
$\langle\,g^{\nu}\rangle$ is contributed by a quantity 
$g^{\nu}_{\alpha\alpha'}\equiv \langle 
RMO_{\alpha}\vert \,{g}^{\nu} \,\vert {RMO}_{\alpha'}\rangle $. Hence, for 
transport problems, the off-diagonal contributions (when $\alpha\neq\alpha'$) 
can be as important as the diagonal ones ($\alpha=\alpha'$). Furthermore, when 
bias $V_b$ is increased, molecular orbitals in the MMM device \cite{Larade}
become less symmetric so that vibrational modes different from the $A_g$ 
symmetry can even contribute to the diagonal matrix elements of the e-p 
coupling. For example, at $V_b= 1$ V, the contribution of orbitals 
${\rm HOMO}-n$ with $n=1,3$ to the scattering states is found to give rise to 
non-zero e-p coupling for vibrational modes with both the $A_{\rm g}$ and  
$B_{\rm 1u}$ symmetries. These results allow us to conclude that e-p coupling 
in MMM devices during current flow ($V_b\neq 0$) is much more complicated than 
that for free molecules.

\begin{figure}[t]
\begin{center}
\leavevmode
\includegraphics[width=12cm]{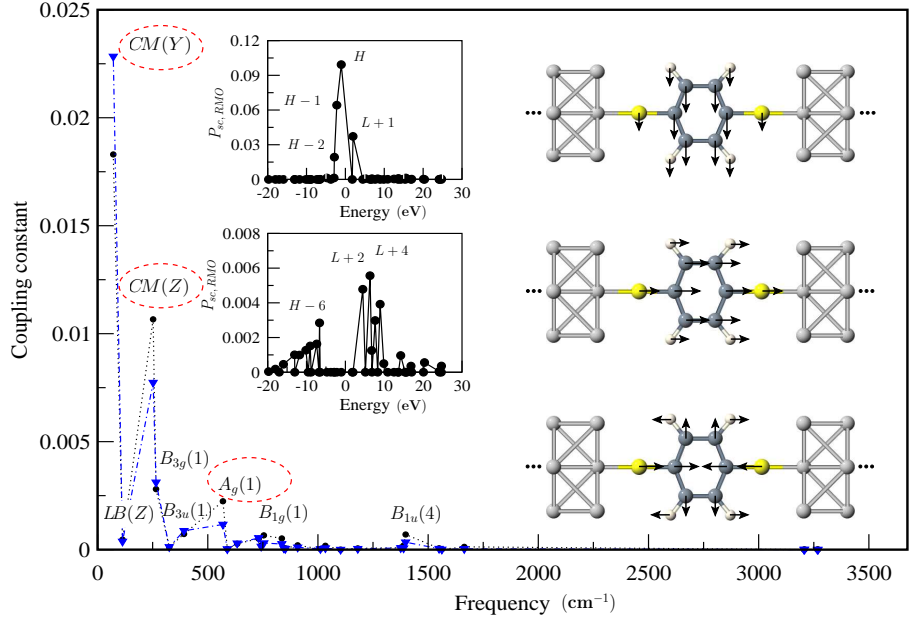}
\end{center}
\caption{Dimensionless e-p coupling constants 
$\lambda_{\nu}^{el}$ (circles) and  $\lambda_{\nu}^{-}$ (triangles) at 
$V_b=1$~ V. Although in principle one should use the $C_{2v}$ point group to 
label the modes, for comparison with Fig. \ref{fig2} we keep the $D_{2h}$ 
labels here. The left insets show projection $P_{sc,\,RMO}$ for two scattering 
states taken at  $E-\varepsilon_{\rm F}^{}=0.4$ eV.  In the right side insets, 
we show eigenvectors of several important modes at $V_b\simeq 1$ V, namely, 
$CM(Y)$ (top), $CM(Z)$ (middle), and $A_{\rm g}(1)$ (bottom).
} 
\label{fig3}
\end{figure}

A most important finding is that at large bias voltage, $V_b \simeq 1$ V, the 
e-p coupling strength changes drastically although the vibrational spectrum is 
only changed by a few percent as mentioned above. Fig. \ref{fig3} plots the
coupling strength at $V_b=1$ V. We observe that contributions to e-p coupling 
are now dominated by a few low-lying modes such as the center-of-mass modes;
the total coupling is a factor of five greater than that at low bias 
(see Fig. \ref{fig2}), and coupling due to individual low-lying modes is also much 
larger. By projecting different scattering states $\Psi_{\rm sc}(E)$ with the 
same energy $E$ onto RMO's as presented above, we found that two patterns of 
$P_{sc,\,RMO}$ occur, as shown in the left insets of Fig. \ref{fig3}. 
This indicates that scattering states inside the transport energy window are 
different from those at low bias. In particular, the pattern of $P_{sc,\,RMO}$ 
in the upper panel is similar to that in Fig. \ref{fig2}, but the new (lower) 
pattern of $P_{sc,\,RMO}$ comes from lower ${\rm HOMO}-n$ and higher 
${\rm LUMO}+n$ RMO's. This leads to a different behavior of e-p couplings at 
large biases for the BDT device. For example, our calculations reveal that the 
peak labeled $CM(Y)$ in $\lambda_{\nu}(\omega_{\nu})$ (see Fig. \ref{fig3}) 
comes from particular off-diagonal matrix elements, 
$\langle L +2\,\vert\,{g}^{\nu} \,\vert\,H-1\rangle$ and  
$\langle L +4\,\vert\,{g}^{\nu} \,\vert\,H -1\rangle$. The peak labeled 
$CM(Z)$, on the other hand, is found to come from diagonal matrix elements, 
{\it e.g.}, $\langle H-1\,\vert\,{g}^{\nu} \,\vert\,H -1\rangle$. 
These findings  also correlate well with the peaks of $P_{sc,\,RMO}$ in the 
left insets of Fig. \ref{fig3}. The total $\lambda_{\rm e-p}$ at 
$V_{b}\simeq 1$ V is found to be $\simeq 0.04$. This enhancement by roughly a 
factor of five from that of $V_b \simeq 0$ is due to the center-of-mass modes 
shown in Fig. \ref{fig3} that can be confirmed by computing 
$\lambda_{\rm e-p}$ without counting these modes. 

\begin{figure}[t]
\begin{center}
\leavevmode
\includegraphics[width=12cm]{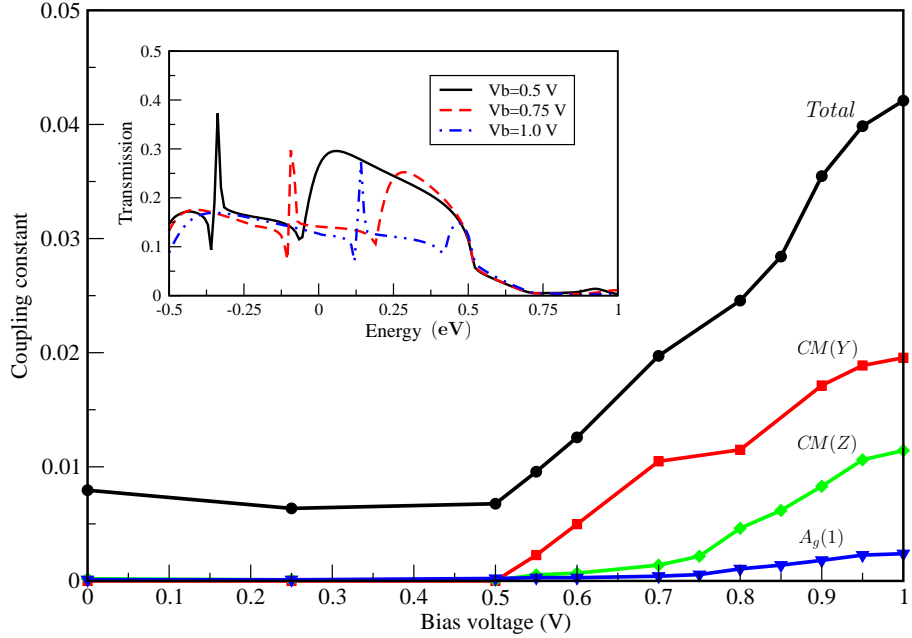}
\end{center}
\caption{  
Dimensionless e-p coupling constants $\lambda_{e-p}^{el}$ (circles) and 
$\lambda_{\nu}^{el}$, for $\nu=CM(Y)$ (squares), $CM(Z)$ (diamonds), and 
$A_{\rm g}(1)$ (triangles), as a function of $V_b$. Inset: transmission 
function $T(E,V_b)$ at $V_b=0.5$ V (full line), $V_b=0.75$ V (dashed line), 
and $V_b=1$ V (dash-dot line). The transport energy window $(\mu_{L}^{},\,\mu_{R}^{})$
lies in the 
positive part of the energy axis, from $E=0$ to $E=\vert eV_{b}\vert $.
} 
\label{fig4}
\end{figure}

Why bias voltage can change e-p coupling so drastically? We found that the 
reason is mainly due to contribution of different scattering states. In
the inset of Fig. \ref{fig4}, we plot transmission coefficient $T(E,V_b)$ 
{\it vs}. electron energy $E$ for several values of $V_b$. Most clearly
shown is that $V_b$ shifts the transmission features toward the transport
window. In particular, a sharp peak (at $E\approx -0.3$ eV) is shifted 
up-wards in energy with the increase of $V_b$. When $V_b > 0.5$ V, the 
``tail'' of this sharp peak starts 
entering  the transport window (between $\mu_{L}^{}=0$ and $\mu_{R}^{}=eV_b$, 
see also Eq. (\ref{lambda1})). If $V_b > 0.75$ V, this peak enters into the
transport window completely. When this happens, the e-p coupling changes 
drastically and the new pattern appears in the projection $P_{sc,\,RMO}$ as 
discussed above. Fig. \ref{fig4} plots the e-p coupling strength 
$\lambda_{\nu}$ for several vibrational modes $\nu$ and the total 
$\lambda_{e-p}$ {\it vs}. $V_{b}$. The curves give a clear indication that 
the e-p coupling is roughly a constant at small biases, but can change 
{\it nonlinearly} as the applied bias voltage is varied. Such a change can 
have deep implications to local heating in the device during nonequilibrium 
charge transport \cite{montgomery2003,jauho2004,diV2003,diV2005}.

In summary, the entire relevant vibrational spectrum of a molecule device 
can be obtained at non-zero bias within the NEGF-DFT formalism where both
vibrational and electronic properties are calculated at equal footing.
For a 1,4-BDT molecular device studied here, low-lying vibrational modes
play an important role in contributing to the e-p coupling strength. The
coupling strength changes drastically as bias voltage is increased due to
participations of new scattering states. For the BDT device, at large bias of 
$V_b\simeq 1$ V, it is the center-of-mass modes which denominate the e-p
coupling, while for small bias $V_b < 0.5$ V, many modes of different
symmetries contribute. The vibrational spectrum also depends on bias, but
for the BDT device this dependence is at a few percent level.

This work is supported by NSERC of Canada, FQRNT of Qu\'{e}bec, and CIAR.

{}


\begin{thebibliography}{}

\bibitem{lorente2000}
N. Lorente and M. Persson, Phys. Rev. Lett. {\bf 85}, 2997 (2000);\newline
N. Lorente, M. Persson, L. J. Lauhon and W. Ho, Phys. Rev. Lett. {\bf 86}, 2593 
(2001).

\bibitem{montgomery2003a}
M. J. Montgomery, J. Hoekstra, T. N. Todorov and A. P. Sutton, 
J. Phys.: Condens. Matter {\bf 15}, 731 (2003).

\bibitem{ho2002}
W. Ho, J. Chem. Phys. {\bf 117}, 11033 (2002).

\bibitem{Reed2004}
W. Wang, T. Lee, I. Kretzschmar and M. A. Reed, Nano Lett. {\bf 4}, 643 (2004).

\bibitem{Mayer05}
Y. Selzer, L. Cai,  M. A. Cabassi, Y. Yao, J. M. Tour, T. S. Mayer,  D. L. Allara,
Nano Lett. {\bf 5}, 61 (2005).

\bibitem{montgomery2003}
M. J. Montgomery and T. N. Todorov, 
{J. Phys.: Condens. Matter} {\bf 15}, 8781 (2003).

\bibitem{ratner2004}
M. Galperin, M. Ratner and A. Nitzan, J. Chem. Phys. {\bf 121}, 11965 (2004).

\bibitem{asai2004}
Y. Asai, Phys. Rev. Lett. {\bf 93}, 246102 (2004). 


\bibitem{jauho2004}
T. Frederiksen, M. Brandbyge, N. Lorente, and A.-P. Jauho, 
Phys. Rev. Lett. {\bf 93}, 256601 (2004).

\bibitem{transiesta}
M. Brandbyge, J.-L. Mozos, P. Ordej\'on, J. Taylor, 
and K. Stokbro, 
{Phys. Rev.} B {\bf 65}, 165401 (2002).


\bibitem{mcdcal}
J. Taylor, H. Guo, and J. Wang, {Phys. Rev.} B {\bf 63},  245407 (2001).

\bibitem{gunnarsson}
O. Gunnarsson, Rev. Mod. Phys. {\bf 69}, 575 (1997).

\bibitem{foot1}
For Hessian matrix, the main difficulty is to obtain the double derivative
of the Hartree and exchange-correlation energies which are functionals of 
charge density that in turn depends on bias voltages. These derivatives can be 
obtained within the NEGF technique. 

\bibitem{foot2}
In the model calculation, the two identical Al leads contact the molecule 
by a distance of $2$ {\AA}. The leads have a finite cross section oriented
in the (100) direction with nine atoms per unit cell and extend to 
$z=\pm \infty$. The vibrational box includes the molecule. A single zeta 
$s,p,d$ basis set is used in the NEGF-DFT self-consistent calculations.

\bibitem{Raman1}
S. H. Cho H. S. Han, D.-J. Jang, K. Kim, and M. S. Kim, J. Phys. Chem. {\bf 99}, 10594 (1995);\newline
S. W. Joo, S. W. Han, and K. Kim, J. Coll. Interf. Sci. {\bf 240}, 391 (2001). 

\bibitem{GTB}
M. Tinkham, {\it Group Theory and Quantum Mechanics} (McGraw-Hill, New York, 
1964).

\bibitem{Larade}
B. Larade,
J. Taylor, Q. R. Zheng, H. Mehrez, P. Pomorski, and H. Guo, 
Phys. Rev. B {\bf 64},  195402 (2001).

\bibitem{diV2003}
Y.-C. Chen, M. Zwolak, and M. Di Ventra, 
{Nano Lett.} {\bf 3}, 1691 (2003); {\it ibid.} {\bf 4}, 1709 (2004).

\bibitem{diV2005}
Z. Yang,  
M. Chsiev, M. Zwolak, Y.-C. Chen, and M. Di Ventra, 
Phys. Rev. B {\bf 71}, 041402 (2005).

\end{thebibliography}
\end{document}